\documentclass[1p,11pt]{elsarticle}
\begin{document}
\begin{frontmatter}

\title {How to implant a causal $\Theta$ function \\
into a tachyon field operator,\\
or why tachyons do not violate causality} 

\author[itep]{Vassili F. Perepelitsa}
\address[itep]{Institute for Theoretical and Experimental Physics, Moscow, Russia\\}
e-mail: vassili.perepelitsa@cern.ch

\begin{abstract}
The postulate of the preferred reference frame 
in which the signal propagation is governed by retarded causality 
is a must for any theory of faster-than-light particles and signals. 
Such a system does exist and is the comoving system of the 
relativistic cosmology. Restrictions imposed by the causality conservation
can be expressed via a causal $\Theta$ function assumed to be acting in both, 
the classical and quantum field theories of tachyons. A Lorentz-covariant 
introduction of this $\Theta$ function, which ensures a causal behaviour of 
real tachyons (asymptotic tachyonic states) preventing the appearance of casual
loops constructed with the use of faster-than-light particles and signals, 
into the tachyon quantum field operators is suggested. 
\end{abstract}

\end{frontmatter}
\pagestyle{plain}

\section{Introduction}
\setcounter{equation}{0}
\renewcommand{\theequation}{1.\arabic{equation}}
During the late fifties and sixties of the last century a possibility
of the introduction of a concept of faster-than-light particles into the 
quantum field theory was considered in several papers \cite{schmid,tan,bds,
wigner,fein,korf}. The particles were called tachyons, from the Greek word
$\tau \alpha \chi \iota \sigma$ meaning $swift$ \cite{fein}.
These considerations have generated a strong critical response based on the
generally accepted principles of causality \cite{newton,roln,parment}, vacuum
stability \cite{nielsen} and unitarity \cite{unita,unita2}. A consensus was 
achieved that within the special relativity and the canonical quantization 
procedure faster-than-light particles are incompatible with those principles. 

Nevertheless, in ref. \cite{ttheor} an approach to a tachyon theory based on 
the requirement of the causality non-violation was suggested. This approach, 
which solves also two other problems of tachyon models mentioned above,
replaces the standard demand to a tachyon theory to be Lorentz-invariant by
a softer condition, requiring the theory to be formulated in a Lorentz-covariant
form, which means {\em an equivalent mathematical description} of tachyon 
behaviour in all reference frames. Then the causality violation by tachyons is 
removed by an introduction of a preferred reference frame in which the events 
of tachyon exchange are ordered by retarded causality, which ensures 
the absence of causal loops in any frame. It was shown in \cite{ttheor}
that the preferred reference frame considered there should be associated with 
the so called comoving frame (see the definition of the latter, for example, 
in \cite{ll1}), being a universal reference frame in which 
our universe is embedded. In particular, the distribution of matter in the 
universe is isotropic in this frame only, the same is true for the relic black
body radiation. The introduction of the preferred frame leads, as a 
straightforward consequence, to the concept of absolute time as the universal 
time acting in the preferred frame. 

The causality protection formula, valid in all inertial frames, was formulated
in \cite{ttheor} as follows:
\begin{equation}
       Pu \geq 0,
\end{equation}
where $P$ is a 4-vector of particle momenta transferring a signal and $u$ is a 
4-velocity of the preferred reference frame with respect to (any particular) 
inertial observer. It is a boundary condition which should be imposed on
solutions of any tachyon equation of motion. 

As a straightforward result, it turns out that the negative energy tachyons, 
which could be used for a construction of the causal loops, cannot appear in 
the preferred reference frame since the eigenvalues of the tachyon Hamiltonian, 
as has been shown in \cite{ttheor} (and will be demonstrated in Sect.~3 of 
this note), are restricted from below, in this frame, by zero value, 
which automatically solves the problem of the tachyon vacuum instability.
The tachyon vacuum in the preferred frame is represented by an infinite
ensemble of zero-energy, but finite-momentum, on-mass-shell tachyons 
propagating isotropically. Thus the space of the preferred frame is spanned by
the continuous background of mass-shell zero-energy tachyons; in some respects 
this is the reincarnation of the ether concept in its tachyonic version. 
Simultaneously it turns out that in any reaction in which tachyons participate
asymptotic ``in" and ``out" tachyonic Fock spaces are unitarily equivalent, 
which removes the unitarity problem.

As ``toy" models, the Lorentz-covariant quantum field models of scalar tachyons 
were considered in ref.~\cite{ttheor} \footnote{It was argued in \cite{ttheor} 
that a realistic model of a tachyon theory should be built upon the 
infinite-dimensional unitary irreducible representations of the Poincar\'{e} 
group (so called ``infinite spin" tachyons). Within the conjecture that 
elementary particles are realizations of the unitary irreducible 
representations of the Poincar\'{e} group the only alternative to the infinite 
spin tachyon models is a scalar tachyon model. However the latter cannot 
represent tachyons at a fundamental level since it possesses several diseases; 
in particular, such a model would lead to the instability of photons 
via their decay to tachyon-antitachyon pairs \cite{ttheor} 
(note that decays of photons to the infinite spin tachyon-antitachyon pairs 
are forbidden by the angular momentum conservation combined with 
kinematic restrictions imposed on this process).

However all individual components of an infinite-dimensional wave equation
must satisfy the Klein-Gordon equation. Therefore the whole argumentation
below concerning the introduction of the causal $\Theta$-function into the 
tachyon field operator and its applications would be also valid for 
infinite-dimensional tachyon models.}. They are based on Lorentz-invariant 
Lagrangians with spontaneously broken Lorentz symmetry, so the 
Lorentz invariance violation appears to be restricted to the tachyon sector 
only, affecting the asymptotic tachyon states 
and leaving the sector of ordinary particles within the Standard Model 
untouched, at least up to presumably small radiative corrections.    
The basic element of those tachyon models is the Lorentz-covariant causal
$\Theta$-function, required by the boundary condition (1.1), 
which ensures the causal behaviour of tachyonic fields and,
subsequently, the other gains of the models. For example, the Hermitian tachyon
field operator with this $\Theta$-function, $\Theta(ku)$, reads as follows:
\begin{equation}
\Phi(x) = \frac{1}{\sqrt{(2\pi)^3}} 
\int{d^4k~\Big{[}a(k)\exp{(-ikx)} + 
a^+(k)\exp{(ikx)}\Big{]}~\delta(k^2+\mu^2)~\Theta(ku)},
\end{equation}
where $k$ is a tachyon four-momentum, $a(k), a^+(k)$ are annihilation and
creation operators with bosonic commutation rules, annihilating or creating 
tachyonic states with 4-momentum $k$, and $\mu$ is a tachyon mass parameter.
As can be seen, the expression (1.2) is explicitly 
Lorentz-covariant. This covariance includes the invariant meaning of the 
creation and annihilation operators defined in the preferred frame;
thus, for example, an annihilation operator $a(k)$ remains an
annihilation operator $a(k^\prime)$ in the boosted frame, even if the zero
component of $k^\prime$ may become negative. 

When calculating the tachyon production probabilities and cross-sections
the confining $\Theta$ functions will accompany the production amplitudes as
factors restricting the reaction phase space, so the expressions for
these probabilities can be displayed as follows:
\begin{equation}
W = \int| M |^2 d\tau \prod_{i} \Theta(k_i u),
\end{equation}
where M is a matrix element of the reaction (which has to be representable in a
Lorentz-invariant form), $d\tau$ is a reaction phase space element, and the
product of $\Theta$ functions includes all free tachyons
(having 4-momenta $k_i$) participating in the reaction.

During discussions of the models above the author was asked to formalize the 
introduction of the causal $\Theta$ function into the tachyon field operator.
This note is a realization of this recommendation on the base of a scalar 
tachyon field model, presenting a  development of the approach to a 
tachyon theory sketched in this Introduction.
 
In formulae used in the note the velocity of light $c$ and the Planck
constant $\hbar$ are taken to be equal to 1.

\section{Causal tachyon field operator}
\setcounter{equation}{0}
\renewcommand{\theequation}{2.\arabic{equation}}
Let us consider a Lorentz-invariant Lagrangian of a free scalar tachyon field
\begin{equation}
L = \frac{1}{2} \int d^3 {\bf x} \Big{[}\dot\Phi^2(x) -
\nabla\Phi(x) \nabla\Phi(x)  + \mu^2 \Phi^2(x) \Big{]}
\end{equation} 
from which the Klein-Gordon equation
with the negative mass-squared term $-m^2 = \mu^2$ follows:
\begin{equation}
\Big{(}\frac{\partial ^2}{\partial t^2} - \partial _i \partial ^i - \mu^2\Big{)} \phi(x) = 0~, ~~~~~~~~i = 1,2,3
\end{equation}
It has well-known solutions in the form of plane waves, so the wave function
of a free particle with a given 4-momentum $k = (\omega,{\bf k})$ must be 
\begin{equation}
 const \times \exp{ -i(\omega t - {\bf kx})} 
\end{equation}
with the dispersion relation
\begin{equation}
\omega = \pm \sqrt{{\bf k}^2 - \mu^2}
\end{equation}
and with the restriction on the particle 3-momentum ${\bf k}$ 
\cite{bds,fein}:
\begin{equation}
|{\bf k}| \geq \mu~.
\end{equation} 
A Fourier representation of the general solution $\phi(x)$,
up to a normalisation factor $1/ \sqrt{(2\pi)^3}$, should be written as 
\begin{equation}
\phi(x)=\int{d^4k~\exp(-ikx)~\delta(k^2+\mu^2)~\phi(k)},
\end{equation}
where $\delta(k^2+\mu^2)$ ensures that 
the field $\phi(x)$ corresponds to particles positioned on the mass shell,
thus validating (2.5), and $\phi(k)$ are Fourier amplitudes. 
A standard problem, appearing at such a decomposition, is related
to the negative sign of the $\omega$ in (2.4) which, being interpreted in a
straightforward way as a particle energy, would lead to a well-known 
problem related to particles with negative energies.  

Our aim is a standard solution of this problem combined with a ``soft" 
(covariant) introduction of a concept of the preferred reference frame 
in the tachyon field model. To do this we introduce two auxiliary 
scalar fields that obey the equation (2.2): 
\begin{equation}
\phi^{(+)}(x) = \frac{1}{2\pi i} \int_{C_+} \phi(x - u\tau) \frac{d\tau}{\tau},
\end{equation}
\begin{equation}
\phi^{(-)}(x) = \frac{1}{2\pi i} \int_{C_+} \phi(x + u\tau) \frac{d\tau}{\tau},
\end{equation}
where $\tau$ is a ``time" parameter (which will be explained below),
$u$, primarily, is some 4-vector of dimension of a 4-velocity, and the contour 
$C_+$ is extended from $-\infty$ to $+\infty$, deformed below the singularity
at $\tau = 0$. These auxiliary fields are similar to those introduced in
\cite{schwinger} in order to define invariantly the plane waves with positive 
and negative ``frequencies". In \cite{schwinger} a 4-vector $\epsilon$, 
formally defined to be timelike and to have a positive time component, was 
used instead of the~$u$. In our consideration the 4-vector $u$ has a physical 
meaning of the 4-velocity of the preferred reference frame with respect to 
the observer, i.e. it is automatically timelike and has a positive time 
component:
\begin{equation}
u_\mu~u^\mu = 1,~~~u_0 = u^0 > 0, ~~~~~~\mu = 0,1,2,3.
\end{equation}
The physical meaning of integrals in (2.7), (2.8) implies the ``collection" of
all virtually allowed field phases (``trajectories") of each individual
field mode dispersed over $\tau$, with a pole at $\tau = 0$.
According to the decomposition (2.6),
\begin{equation}
\phi^{(+)}(x) = \int {d^4k~\exp(-ikx)~\delta(k^2+\mu^2)~ 
\phi^{(+)}(k)~\frac{1}{2\pi i} 
\int_{C^+} \exp( i~k u~\tau){\frac{d\tau}{\tau}}},
\end{equation}
\begin{equation}
\phi^{(-)}(x) = \int {d^4k~\exp(-ikx)~\delta(k^2+\mu^2)~
\phi^{(-)}(k)~\frac{1}{2\pi i} 
\int_{C^+} \exp(-i~k u~\tau){\frac{d\tau}{\tau}}}.
\end{equation}
Noting that
\begin{equation}
\frac{1}{2\pi i} \int_{C^+} \exp( i~k u~\tau){\frac{d\tau}{\tau}} =
\cases{1~~~& if~$ku  >  0$ \cr
       0~~~& if $ku  <  0$ \cr }
\end{equation}
and        
\begin{equation}
\frac{1}{2\pi i} \int_{C^+} \exp(-i~k u~\tau){\frac{d\tau}{\tau}} =
\cases{1~~~& if~$ku <   0$ \cr
       0~~~& if $ku >   0$ \cr }
\end{equation}
the equations (2.10), (2.11) can be rewritten as
\begin{equation}
\phi^{(+)}(x) = \int {d^4k~\exp(-ikx)~\delta(k^2+\mu^2)~ 
\phi^{(+)}(k)~\Theta(ku)},
\end{equation}
\begin{equation}
\phi^{(-)}(x) = \int {d^4k~\exp(-ikx)~\delta(k^2+\mu^2)~
\phi^{(-)}(k)~\Theta(-ku)}.
\end{equation}

Let us consider these equations in the preferred reference frame, where
$u = (1,0,0,0)$:
\begin{equation}
\phi^{(+)}(x) = \int {d^4k~\exp(-ikx)~\delta(k^2+\mu^2)~ 
\phi^{(+)}(k)~\Theta(\omega)},
\end{equation}
\begin{equation}
\phi^{(-)}(x) = \int {d^4k~\exp(-ikx)~\delta(k^2+\mu^2)~
\phi^{(-)}(k)~\Theta(-\omega)}.
\end{equation}
One can see that in the preferred frame 
$\phi^{(+)}(x)$ differs from zero only at positive ``frequencies" 
(the upper sign for $\omega$ in (2.4)), while $\phi^{(-)}(x)$ differs from zero 
only at negative ``frequencies" (the lower sign for $\omega$ in (2.4)).
Introducing the definition $E = \omega$ in (2.16) we can return to the 
covariant expression (2.14) for  $\phi^{(+)}(x)$, where $k$ is redefined as
\begin{equation}
k=(E,{\bf k}).
\end{equation}
Analogously, introducing the definition $E = -\omega > 0$ in (2.17) and using 
the identity
\begin{equation}
\int d^3{\bf k}\exp(i{\bf kx})
\int_{C^+} \exp[i(E u_o + {\bf ku})\tau]{\frac{d\tau}{\tau}}=
\int d^3{\bf k}\exp(-i{\bf kx})
\int_{C^+} \exp[i(E u_o - {\bf ku})\tau]{\frac{d\tau}{\tau}}
\end{equation}
we can rewrite (2.15) as
\begin{equation}
\phi^{(-)}(x) = \int {d^4k~\exp(ikx)~\delta(k^2+\mu^2)~
\phi^{(-)}(k)~\Theta(ku)}
\end{equation} 
with the definition (2.18) for $k$.
    
Thus we have arrived at a result that in both functions, $\phi^{(+)}(x)$ and 
$\phi^{(-)}(x)$, the variable $E$, which will be referred to in what follows 
as a particle energy, is, by the construction, always positive 
in the preferred reference frame. Keeping this in mind, we can write 
a general solution of (2.2) in an arbitrary frame in the form
\begin{equation}
\Phi(x) = \int {d^4k~[\exp(-ikx)~a(k) + \exp(ikx)~a^+(k)]
~\delta(k^2+\mu^2)~\Theta(ku)},
\end{equation}
the amplitudes $\phi^{(+)}(k),~\phi^{(-)}(k)$ in the previous expressions 
for $\phi^{(+)}(x),~\phi^{(-)}(x)$ being replaced, by a procedure of 
second quantization, by annihilation
and creation operators $a(k),~a^+(k)$, respectively, annihilating and creating 
states with 4-momentum $k$. As mentioned in the Introduction,
the meanings of the operators $a(k), a^+(k)$ to be
the annihilation and creation operators are conserved in an arbitrary frame,
even if the energy component of a 4-vector $k^\prime$ in a boosted frame
(i.e. $E^\prime$) can become negative under a suitable proper Lorentz 
transformation. 

\section{Chronology protection agency at work}
\setcounter{equation}{0}
\renewcommand{\theequation}{3.\arabic{equation}}
Let us consider how the causal $\Theta$-function prevents the violation of
the principle of causality by the use of tachyons.  

The principle of causality states that any (tachyon) theoretical model 
should not admit an appearance of causal loops, i.e. the possibility of
sending by an observer signals to its own past. However within special
relativity such a possibility does exist for faster-than-light signals, 
as proved by R.~C.~Tolman in 1917 (long before the tachyon hypothesis was 
formulated, \cite{tolman}, see also \cite{moller,bohm}). 
The introduction of the concept of the preferred reference frame, 
in which tachyon interactions are ordered by retarded causality, 
allows to destroy this possibility, as mentioned in the Introduction. 
Now we shall trace how this destruction occurs exploiting a concrete (scalar) 
tachyon model.

Using (2.1) and (2.21), the latter following from the former after an
introduction of the concept of the preferred reference frame, one can 
proceed, integrating (2.21) over $k_0$ and after expressing canonical
annihilation and creation operators $a_{{\bf k}}, a^+_{{\bf k}}$, annihilating
or creating tachyonic states with 3-momentum ${\bf k}$, via $a(k), a^+(k)$
\begin{equation}
a_{{\bf k}} = a(k)~\Theta(ku)/\sqrt{2(ku)},
\label{eq:301}
\end{equation}
\begin{equation}
a^+_{{\bf k}} = a^+(k)~\Theta(ku)/\sqrt{2(ku)},
\label{eq:302}
\end{equation}
with the denominators included to ensure a proper covariant normalisation of a
single-tachyon wave function, to obtain an expression for the tachyon field
operator in the form
\begin{equation}
\Phi(t,{\bf x}) = \int_{|{\bf k}| > \mu,E > {\bf ku}}
{\frac{d^3{\bf k}}{2 E}} \sqrt{\frac{2(E - 
{\bf ku})}{(2\pi)^3 \sqrt{1-u^2}}}
~\Big{[}a_{{\bf k}}\exp{(-i E t + i{\bf kx})} + 
a^+_{{\bf k}}\exp{(i E t - i{\bf kx})}\Big{]}.
\end{equation}
Returning to (2.21) and requiring the field to obey the translational 
invariance the following equation should hold:
\begin{equation}
 [P_\mu, \Phi(x)] = -i \partial_\mu \Phi(x), 
\end{equation}
where $P_\mu$ is an operator of a 4-momentum of the field. Its solution
for  $P_\mu$ is: 
\begin{equation}
P_\mu=\frac{1}{2}\int{\frac{d^4k}{(2\pi)^3}~k_\mu~[a^+(k)a(k)+a(k)a^+(k)]~\delta(k^2+\mu^2)~\Theta(ku)}
\end{equation}
when choosing the bosonic commutation relations for $a, a^+$ operators:
\begin{equation}
[a(k), a(k^\prime)] = 0, ~~~[a^+(k), a^+(k^\prime)] = 0.
\end{equation}
\begin{equation}
[a(k), a^+(k^\prime)]~\delta(k^2 +\mu^2)~\delta(k^{\prime 2}+\mu^2)~
\Theta(ku)~\Theta(k^\prime u) = 
\delta^4 (k - k^\prime)~\delta(k^2 +\mu^2)~\Theta(ku).
\end{equation}
In particular, the field Hamiltonian is
\begin{equation}
H \equiv P_0 = \frac{1}{2}\int{\frac{d^4k}{(2\pi)^3}~k_0~[a^+(k)a(k)+a(k)a^+(k)]~\delta(k^2+\mu^2)~\Theta(ku)}.
\end{equation}
Passing again to a 3-dimensional integral with the use of \ref{eq:301},
\ref{eq:302}, followed by dropping as usually the infinite $c$-number related 
to zero-point oscillations, we arrive at the Hamiltonian    
\begin{equation}
H=\int_{|{\bf k}| > \mu, E>{\bf ku}}
{\frac{d^3{\bf k}}{(2\pi)^3}\frac{E-{\bf ku}}{\sqrt{1-{\bf u}^2}}
~a^+_{{\bf k}} a_{{\bf k}}},
\label{eq:303}
\end{equation}
whose eigenvalues are bounded from below. This results, in particular,
in a possibility of the standard operator definition of the invariant 
vacuum state $|0>$ via the annihilation operators $a(k)$, $a(k)|0> = 0$ 
for all $k$ such that $|{\bf k}| > \mu$, with the vacuum
energy boundaries to be defined by the tachyon vacuum gauge
\begin{equation}
     Pu = 0.
\label{eq:304}
\end{equation} 
For example, in the frames moving with respect to the preferred frame the 
energy boundaries of the tachyon vacuum are given by expressions
\begin{equation}
E_0^+ = \frac{\mu |{\bf u}|}{\sqrt{1 - {\bf u^2}}}
\label{eq:305}
\end{equation}
for the direction of motion of the preferred frame coinciding with that of the 
tachyon velocity ${\bf v}$, and by 
\begin{equation}
E_0^- = -\frac{\mu |{\bf u}|}{\sqrt{1 - {\bf u^2}}}.
\label{eq:306}
\end{equation}  
for the opposite direction. Thus the tachyon 
vacuum energy boundaries are rotationally invariant in the preferred reference 
frame only. In this frame
\begin{equation}
H = \int_{|{\bf k}| > \mu,E>0}
{\frac{d^3{\bf k}}{(2\pi)^3}~E ~a^+_{{\bf k}} a_{{\bf k}}}
\end{equation}
having non-negative eigenvalues.

Just the equation (\ref{eq:305}) prevents the construction of causal loops
with the use of tachyons: to build such a loop one needs to send tachyons
having velocities $|{\bf v}| > 1/|{\bf u}|$ along the direction of motion
of the preferred reference frame ${\bf u}$, 
but such velocities, as can be easily seen, would correspond to 
tachyon energies below the allowed energy limit in this 
direction (\ref{eq:305}), i.e. they are forbidden.  
One can say that acasual tachyons get confined within the tachyon vacuum.

It is interesting to note that formula (1.1) works in the case of ordinary
particles also, destroying the possibility of having negative energies of
these particles. This suggests the idea that this formula has a general 
application and can be considered as a realization of ``the chronology 
protection agency", the term being primarily introduced by 
S.~W.~Hawking~\cite{hawking} to protect the causality in some general 
relativity applications. Thus, the great efforts undertaken 
to resolve the problem of particle negative energies appearing in relativistic
quantum theory may be considered, formulating an alternative point of view, 
as equivalent to the introduction of the concept of the preferred reference 
frame (and hence ``absolute time") in the philosophy of that theory, 
which trivially solves the problem, analogously to the prescriptions (2.7), 
(2.8) for a scalar tachyon field.
 
To make this story complete let us remark that the need of an introduction of
the concept of the preferred reference frame in the quantum theory
in order to ensure the conservation of causality was noticed long time ago by
P.~H.~Eberhard in \cite{eber}, the idea being shared also by J.~S.~Bell 
\cite{bell1,bell2}. 

\section{Conclusion}
The introduction of the causal $\Theta$ function $\Theta(ku)$ into the
tachyon field operator, where $k$ is a tachyon 4-momentum 
and $u$ is a 4-velocity of the preferred reference frame
in which tachyon interactions are ordered by retarded causality, aimed at
solving several serious problems of a theory of faster-than-light
particles, including the violation of causality by tachyons, is formalized 
in this note within a Lorentz-covariant approach.

\section*{Acknowledgements}
I express my gratitude to Prof. O.~V.~Kancheli for pointing out the problem  
of introducing the causal $\Theta$ function into the tachyon field operator, 
to Profs.~K.~G.~Boreskov, F.~S.~Dzheparov and S.~M.~Sibiryakov for stimulating 
discussions, and to Dr.~B.~R.~French for the critical reading of the manuscript.


\end{document}